# Energy Level Alignment in Ternary Organic Solar Cells

*Vincent Lami[1], Yvonne J. Hofstetter[1,2], Julian F. Butscher[1], Yana Vaynzof[1,2]\**

[1] Kirchhoff Institute for Physics and Centre for Advanced Materials, Ruprecht-Karls-Universität Heidelberg, Im Neuenheimer Feld 227, 69120 Heidelberg, Germany

[2] Integrated Centre for Applied Physics and Photonic Materials and Centre for Advancing Electronics Dresden (cfaed), Technical University of Dresden, Nöthnitzer Str. 61, 01187 Dresden, Germany

**Corresponding Author**

*Yana Vaynzof, e-mail: yana.vaynzof@tu-dresden.de

## ABSTRACT

Ternary organic solar cells (TOSC) are currently under intensive investigation, recently reaching a record efficiency of 17.1%. The origin of the device open-circuit voltage ($V_{OC}$), already a multifaceted issue in binary OSC, is even more complex in TOSCs. Herein, we investigate two ternary systems with one donor (D) and two acceptor materials (A1, A2) including fullerene and non-fullerene acceptors. By varying the ratio between the two acceptors, we find the $V_{OC}$ to be gradually tuned between those of the two binary systems, D:A1 and D:A2. To investigate the origin of this change, we employ ultra-violet photoemission spectroscopy (UPS) depth profiling, which is used to estimate the photovoltaic gap in the ternary systems. Our results reveal an excellent agreement between the estimated photovoltaic gap and the $V_{OC}$ for all mixing ratios, suggesting that the energetic alignment between the blend components varies depending on the ratio D:A1:A2. Furthermore, our results indicate that the sum of radiative and non-radiative losses in these ternary systems is independent of the blend composition. Finally, we demonstrate the superiority of UPS over X-ray photoemission spectroscopy (XPS) depth profiling in resolving compositional profiles for material combinations with very similar chemical, but dissimilar electronic structures, as common in TOSCs.

**TOC GRAPHICS**

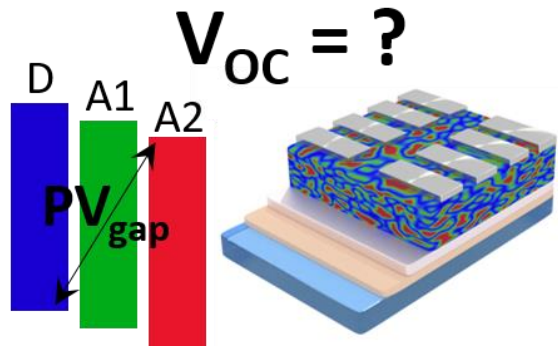

**INTRODUCTION**

Solution-processability at low temperatures, semi-transparency, flexibility, and low weight make organic solar cells (OSCs) attractive candidates for future applications.[1,2] Both the invention of the bulk heterojunction (BHJ) concept for the active layer morphology,[3,4] and the introduction of non-fullerene acceptors (NFAs)[5–11] resulted in tremendous improvements in device performance, yielding record power conversion efficiency (PCE) values exceeding 16.5 % for single-junction devices.[12][13] However, the incomplete absorption of the solar spectrum and non-avoidable radiative as well as non-radiative recombination losses are still limiting important device parameters, in particular the open-circuit voltage ($V_{OC}$), and consequently the overall PCE.[14]

In recent years, novel device architectures and other optimization strategies have been proposed to design better performing OSCs and to overcome their limitations. One of these strategies focuses on tandem devices with multiple active sub-layers enhancing the amount of absorbed light by complementary light absorption.[15–17] The fabrication of such stacked layers is quite challenging,[18] prompting the development of ternary organic solar cells (TOSCs) which provide a promising alternative. Compared to standard binary BHJ solar cells, TOSCs employ an active layer consisting of not only two but three materials, of which one can be a donor (D) and two can be acceptors (A) or the other way round.[14,19–23] Incorporating an additional, third component offers the potential to broaden the absorption window and to enhance the charge mobility, thereby increasing both the short-circuit current density ($J_{SC}$)[20,24,25] and fill factor (FF)[21,26,27]. Furthermore, optimizing material mixing ratios has been suggested to minimize non-radiative loss channels which in turn enhances both the FF and the $V_{OC}$.[28] To date, TOSC efficiencies exceeding 14 % with the current record being 17.1 % have been demonstrated,[29][13] motivating further research aimed at the enhancement of device efficiency and investigation of TOSC device physics and photophysics.

Many aspects of ternary blends, including recombination dynamics, the location of the third component in the blend, energetic alignment, and the tunability of the $V_{OC}$ by changing the mixing ratio of blend components, are still not entirely understood.[19,30–35] To address the latter, several models, such as the energy cascade,[36,37] parallel junction,[38] alloy,[30,39] and the Gaussian disorder model[14,23] have been developed with the goal of explaining the tunable character of the $V_{OC}$. The energy cascade model predicts values close to that of the lower $V_{OC}$ binary device. Alternatively, in the parallel junction model, the active layer is considered as a combination of two independent sub-active layers made up of the two possible D:A combinations, yielding a $V_{OC}$ value that lies between those of the two corresponding binary blends.[38] Despite this, the parallel junction model has also been applied to ternary systems in which the ternary blend $V_{OC}$ is limited by that of the binary system with the lower $V_{OC}$.[40] The Gaussian disorder model and alloy model predict a tunable $V_{OC}$ in ternary devices, however, both are highly dependent on material miscibility. The alloy model, on the one hand, predicts that the $V_{OC}$ originates from the interface band gap, but fails to explain the $V_{OC}$ evolution for fullerene based devices.[41,42] The Gaussian disorder model, on the other hand, includes the disorder driven broadening of the highest occupied molecular orbital (HOMO) and the lowest unoccupied molecular orbital (LUMO) levels and takes quasi Fermi level splitting into account, resulting in an agreement with experimental values.[14,23] Recently, it has been recently suggested by Gasparini *et al.*[18] that the $V_{OC}$ value is mainly determined by the lowest of the charge transfer (CT) state energies in the ternary active layer. This was proposed to occur as a consequence of the higher energy CT state relaxation to the lowest energy CT state prior to charge separation.

While energy level diagrams of TOSCs are routinely reported in literature,[13,28,32,34,39,40,43] it is important to note that these diagrams are not directly measured on the device active layer and thus do not represent the actual energetic landscape of ternary systems. As a result, experimental investigation of the correlation between the changes in the device open-circuit voltage and the evolution of energetics in ternary blends has not been performed to date. Herein, we apply our recently developed ultraviolet photoemission spectroscopy (UPS) depth profiling technique[44] to TOSC systems in order to probe their vertical energetic and compositional landscape. We investigate two TOSC systems and quantify the relationship between the measured photovoltaic gap and open-circuit voltage. The first system we examine is based on poly[(2,6-(4,8-bis(5-(2-ethylhexyl-3-chloro)thiophen-2-yl)-benzo[1,2-b:4,5-b']dithiophene))-alt-(5,5-(1',3'-di-2-thienyl-5',7'-bis(2-ethylhexyl)benzo[1',2'-c:4',5'-c']dithiophene-4,8-dione)] (PBDB-T-2Cl) in combination with [6,6]-Phenyl-C71-butyric acid methyl ester (PC$_{70}$BM) and 3,9-bis(2-methylene-((3-(1,1-dicyanomethylene)-6,7-difluoro)-indanone))-5,5,11,11-tetrakis(4-hexylphenyl)-dithieno[2,3-d:2',3'-d']-s-indaceno[1,2-b:5,6-b']dithiophene (ITIC-2F)). In this ternary system, the non-fullerene acceptor ITIC-2F complements the absorption of PC$_{70}$BM, offering the opportunity to increase the short-circuit current. The second system we investigate is based on poly[[4,8-bis[(2-ethylhexyl)oxy]benzo[1,2-b:4,5-b']dithiophene-2,6-diyl][3-fluoro-2-[(2-ethylhexyl)carbonyl]thieno[3,4-b]thiophenediyl]] (PTB7) together with two fullerene acceptors PC$_{70}$BM and indene-C60 bisadduct (ICBA). This system, originally investigated by Cheng et al,[43] has been proposed to operate via electron cascade mechanism. The chemical structures of the donor polymers and the acceptor molecules of the two ternary systems are shown in **Fig. 1a** and **1c** with a schematic of the device architecture shown in **Fig. 1b**. For both systems, we characterize the photovoltaic performance of the TOSCs and find an excellent agreement between the measured $V_{OC}$ and the smaller photovoltaic gap probed by UPS depth profiling. Beyond energetic information, we find that UPS depth profiling serves as a powerful tool to examine vertical compositional profiles in ternary systems. By comparing the results to those obtained by X-ray photoemission spectroscopy (XPS) depth profiling, we find that UPS outperforms XPS depth profiling due to its ability to distinguish between similar chemical structures such as $PC_{70}BM$ and ICBA, solely due to their different electronic structure. These results highlight the importance of correctly characterizing the vertical energetic and compositional profiles of ternary blends in order to explain and/or predict their photovoltaic device performance and demonstrate the efficacy of UPS depth profiling in attaining this goal.

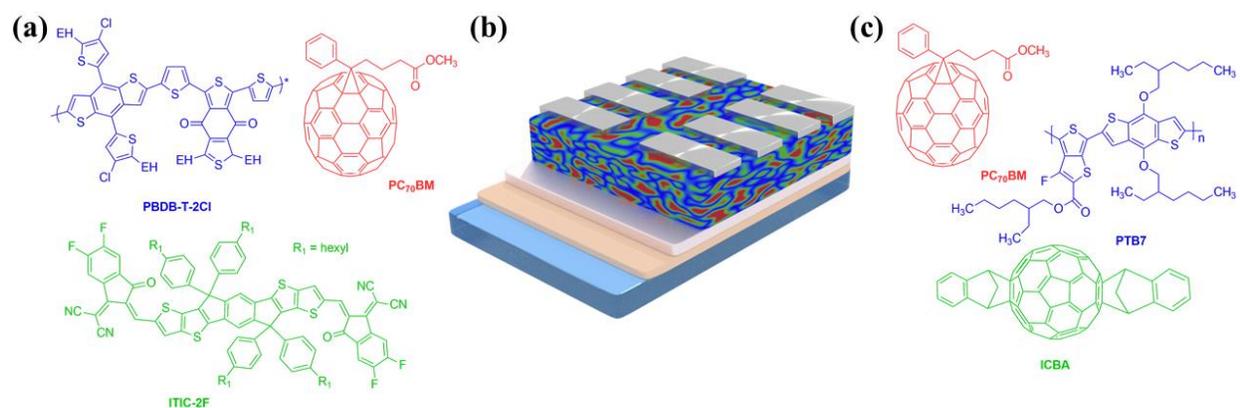

**Figure 1**. Chemical structure of the tested materials in (a) and (c) and a schematic of the device architecture in (b).

**RESULTS**

We fabricated inverted architecture TOSCs with varying A1:A2 weight ratios, while keeping the donor weight ratio constant. The current density-voltage (J-V) characteristics of the best devices are shown in **Fig. 2** with corresponding external quantum efficiency (EQE) curves and photovoltaic performance parameters shown in **Suppl. Fig. 1** and **2**, respectively. We note that for both system, we chose to fabricate devices with relatively thin active layers (50-60 nm), which lower the short-circuit current and overall efficiency, but allow for a more reliable characterization by UPS depth profiling. Importantly, this does not influence the trend in the open-circuit voltage of the devices as evidenced by thicker devices that showed nearly identical $V_{OC}$ values (not shown here).

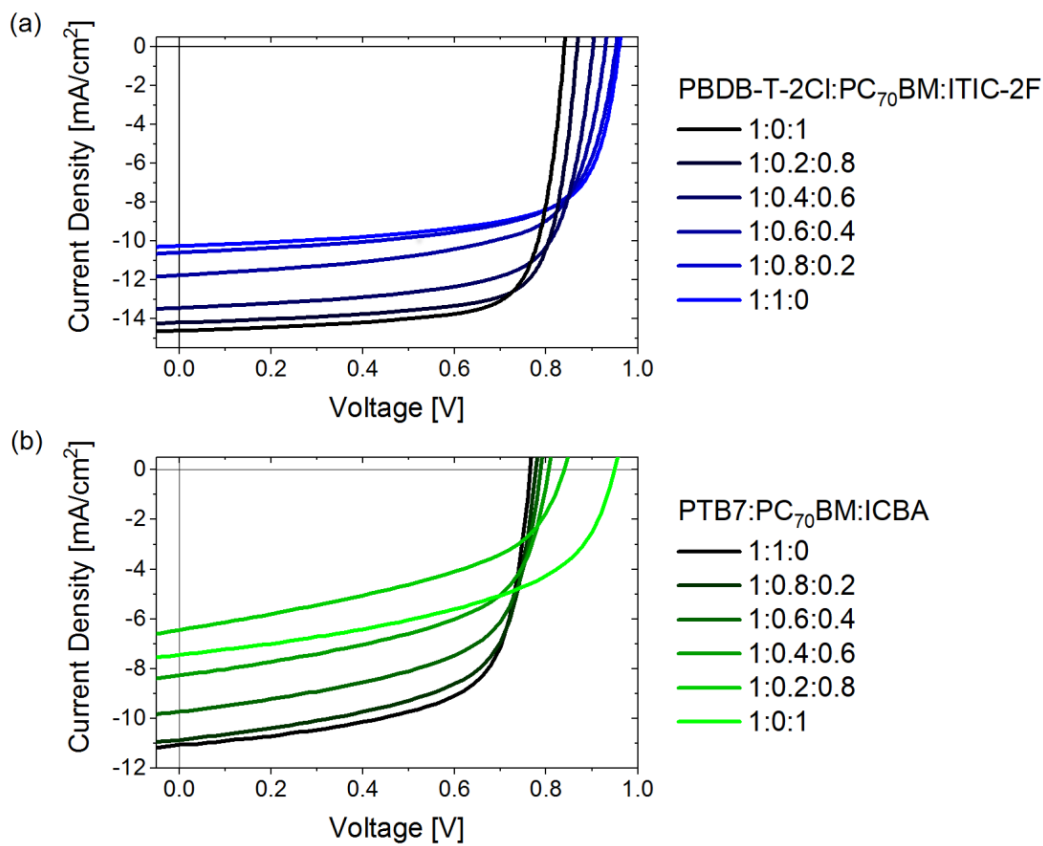

**Figure 2**. Hero J-V curves for both investigated systems: (a) PBDB-T-2Cl:PC$_{70}$BM:ITIC-2F and (b) PTB7:PC$_{70}$BM:ICBA for different acceptor:acceptor mixing ratios. Corresponding EQE curves and J-V parameters are summarized in **Suppl. Fig. 1** and **2**, respectively.

In the first system, PBDB-T-2Cl:PC$_{70}$BM:ITIC-2F, the introduction of increasing amounts of ITIC-2F results in an enhancement of the $J_{SC}$ due to the additional absorption provided by the NFA as is confirmed by the change in the EQE spectra (**Suppl. Fig. 1**) between 700 nm and 800 nm. This boost in $J_{SC}$ is accompanied by a rise in FF and a decrease in $V_{OC}$, which gradually changes from 0.95 V for PBDB-T-2Cl:PC$_{70}$BM to 0.83 V for PBDB-T-2Cl:ITIC-2F. These changes lead to an increase in the PCE from 6.7 % to 9.2 %, as summarized in **Suppl. Fig. 2**. Devices with a PC$_{70}$BM weight ratio of 20 % performed best, achieving a maximum PCE of 9.3 %.

In the second system, PTB7:PC$_{70}$BM:ICBA, changing the ratio of the two fullerene acceptors results in a fairly similar trend: while the $V_{OC}$ is increased with increasing amounts of ICBA, the $J_{SC}$ and FF are decreased. It is interesting to note that while the $V_{OC}$ of both systems is gradually tuned in both cases, the functionality of this change is different for the two cases. In the case of PBDB-T-2Cl:PC$_{70}$BM:ITIC-2F, the $V_{OC}$ tuned by 0.12 V relatively gradually, in a quasi-linear fashion. On the other hand, in PTB7:PC$_{70}$BM:ICBA the $V_{OC}$ is only very slightly increased until an ICBA ratio of 60%, with most of the change in $V_{OC}$ occurring at higher amounts of ICBA.

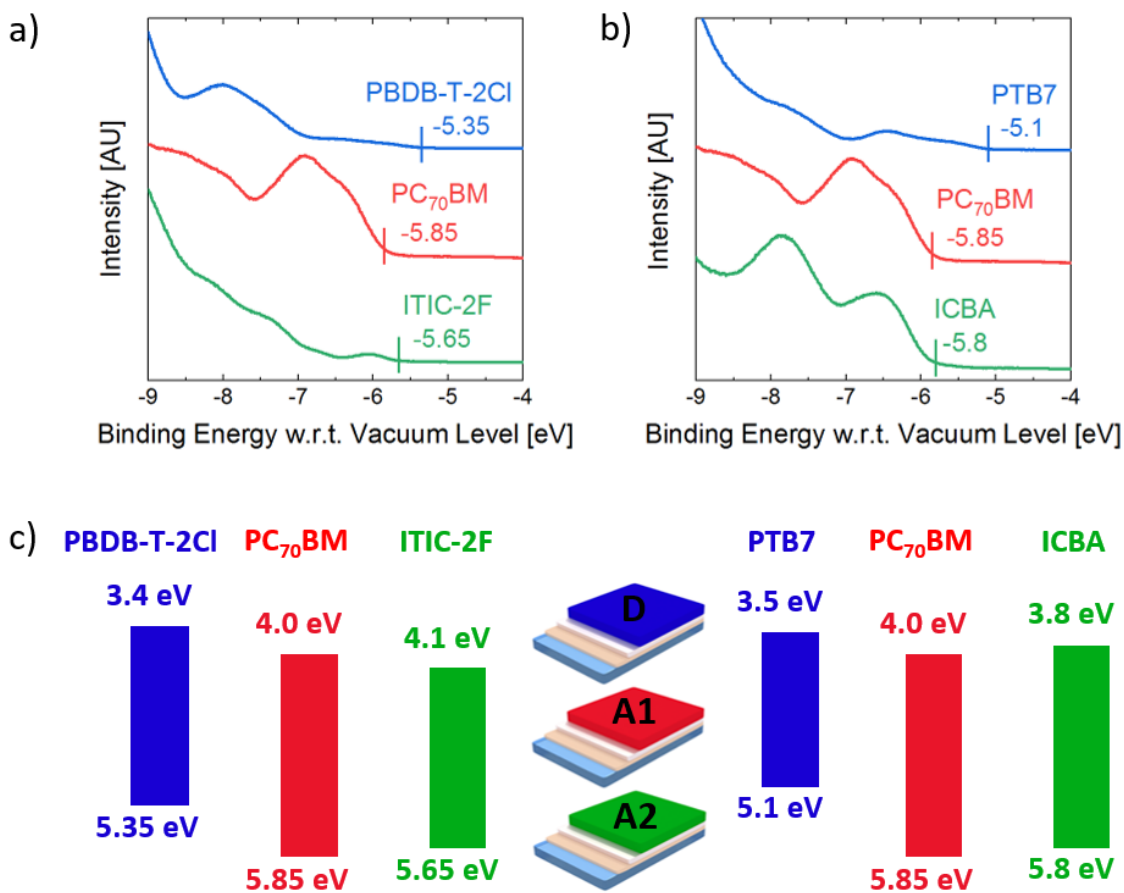

**Figure 3**. UPS spectra of the components of (a) PBDB-T-2Cl:PC70BM:ITIC-2F and (b) PTB7:PC70BM:ICBA ternary systems. The HOMO onset and corresponding energy value (i.e. the ionization potential) are indicated for each spectrum. (c) energetic diagrams showing the individually measured HOMO and estimated LUMO levels for both sets of materials tested in the ternary systems, without taking into account interfacial effects.

To investigate the relationship between the energetic landscape and the $V_{OC}$ in these systems, we performed ultra-violet photoemission spectroscopy measurements on each of the pristine blend components (**Fig. 3a** and **3b**), which are used to extract their ionization potential by extrapolating the HOMO level onset with respect to vacuum level. By subtracting the optical bandgap of each material, we are able to estimate their electron affinity as well. The energy level diagrams are summarized in **Fig. 3c**. From these energy level diagrams, it is possible to estimate the photovoltaic gaps given by the difference between the ionisation potential of the donor and the electron affinity of each of the two acceptor materials. The energetic difference between the photovoltaic gaps of the two binary systems PBDB-T-2Cl:PC$_{70}$BM and PBDB-T-2Cl:ITIC-2F is approximately 0.1 eV, in good agreement with the difference between their $V_{OC}$. Similarly, the photovoltaic bandgaps of PTB7:PC$_{70}$BM and PTB7:ICBA differ by approximately 0.2 eV, explaining the increase in $V_{OC}$ by 0.18 V. How these energy level diagrams can be used to predict the $V_{OC}$ of the ternary blends, however, is unclear. It has been suggested by Gasparini *et al*[17] that the $V_{OC}$ is predominantly determined by the lowest CT state energy in the active layer, *i.e.* by the lower one of the two photovoltaic gaps. Considering that in such an energy level diagram the lower photovoltaic gap remains unchanged for the different mixing ratios, one might mistakenly assign the increase in $V_{OC}$ to, for example, a decrease in radiative and non-radiative losses. However, it is important to remember that such energy level diagrams are constructed from UPS measurements performed on the pure material films, which do not account for interfacial effects that occur when the materials come in contact. Additionally, UPS is a surface sensitive measurement technique and does not provide any information about the bulk material. Consequently, energy level diagrams as shown in **Fig. 3** are not suitable for the interpretation of device physics in ternary systems and do not correctly represent their energetic landscape.

In order to gain a more realistic picture of the vertical energetic landscape throughout ternary blend films, we applied high resolution UPS depth profiling,[44] which allows us to estimate the photovoltaic gap at each depth in both ternary systems for all mixing ratios. In short, this method combines UPS with gas cluster ion etching, allowing us to acquire a UPS spectrum for every depth of the active layer. We then fit the valence band spectra of the ternary blend with a linear combination of the three valence band spectra of the pure materials (**Fig. 3a and b**) to determine the energetic position of the HOMO of each of the three materials in the blend. Then, we are able to approximate their electron affinities by subtracting the optical bandgap of each material and using these values, to estimate the lower photovoltaic gap at each depth. We would like to point out that, while being a destructive method, UPS depth profiling utilizes an argon gas cluster ion etching beam which has been shown to minimize etching damage in organic materials and to preserve the electronic and compositional structure of organic materials.[44,45]

To compare the values found by UPS depth profiling to the measured $V_{OC}$, we average the smaller photovoltaic gap over the entire active layer thickness. We also estimate the error in the photovoltaic gap, which is influenced by the experimental error of the UPS measurement, errors introduced by fitting as well as the result of averaging over the entire active layer. **Fig. 4** shows the comparison of the smaller photovoltaic gap and the corresponding device $V_{OC}$ as a function of the $PC_{70}BM$ weight ratio for the PBDB-T-2Cl:$PC_{70}BM$:ITIC-2F system (left) and the ICBA weight ratio for the PTB7:$PC_{70}BM$:ICBA system (right). We note that although the two y-axes are shifted with respect to each other, they have the same scaling, which shows that the increase in both values follows the same functionality. These results confirm that the energetic alignment between the ternary blend components changes depending on their mixing ratio, which can be attributed to changes in active layer microstructure and disorder.[30]

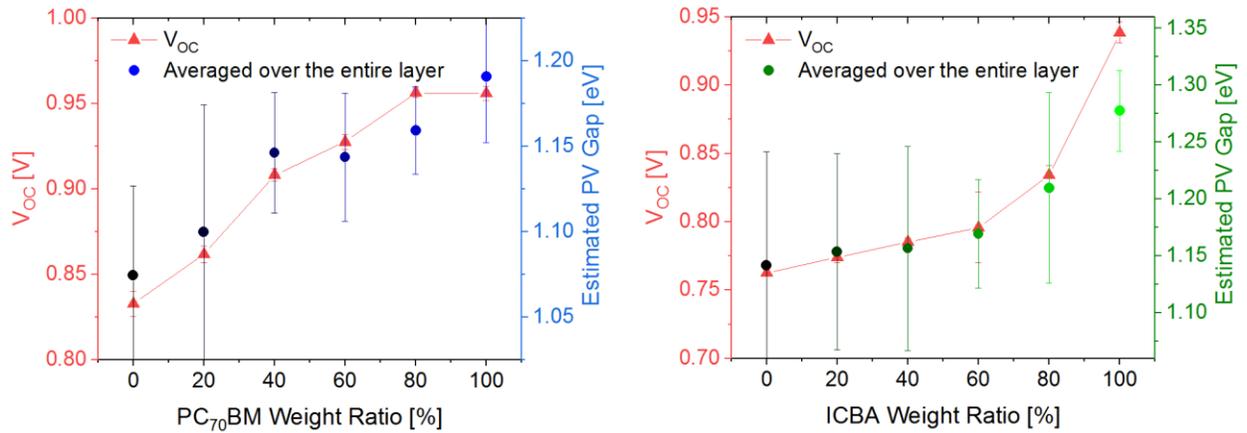

**Figure 4.** Estimated photovoltaic gaps and corresponding device $V_{OC}$s plotted with the same scaling for both investigated ternary systems, PBDB-T-2Cl:PC$_{70}$BM:ITIC-2F on the left and PTB7:PC$_{70}$BM:ICBA on the right for all investigated material mixing ratios..

To quantify the agreement between the values, we plotted the $V_{OC}$ as a function of the photovoltaic gap as shown in **Suppl. Fig. 3**. A linear fit of the values resulted in a slope of $(1.16 \pm 0.17)$ V/eV and $(1.24 \pm 0.08)$ V/eV for the PBDB-T-2Cl:PC$_{70}$BM:ITIC-2F and PTB7:PC$_{70}$BM:ICBA system, respectively. The slightly larger slope of the latter system is partly a consequence of the high $V_{OC}$ of the PTB7:ICBA system, which exhibits lower losses as compared to PTB7:PC$_{70}$BM. Nevertheless, such an excellent correlation between the measured $V_{OC}$ and the photovoltaic gap (a slope of nearly unity) suggests that the sum of radiative and non-radiative recombination losses does not change significantly when varying the mixing ratio of the acceptors. The sum of these losses, which is given by the difference between photovoltaic gap and the $V_{OC}$, is represented by the x-axis intercept in **Suppl. Fig. 3**. In the PBDB-T-2Cl:PC$_{70}$BM:ITIC-2F system this intercept equals to $0.35 \pm 0.12$ eV, significantly lower than the $0.53 \pm 0.04$ eV measured for the PTB7:PC$_{70}$BM:ICBA system. We note, that compared with other state-of-the-art systems, PBDB-T-2Cl:PC$_{70}$BM:ITIC-2F-based ternaries exhibit very low recombination losses.[44]

In addition to energetic depth profiles, UPS provides valuable information concerning the compositional distribution of materials in ternary blends. We compare the compositional profiles as probed by both XPS and UPS depth profiling in **Fig. 5**. All contributing materials are plotted separately for more clarity. In both material systems, XPS depth profiling yields widely varying compositional profiles (on the order of 20%) which probably originate from the compositional similarity between the ternary blend components, resulting in relatively large errors when attempting to distinguish between them. This is particularly difficult in the case of PTB7:PC$_{70}$BM:ICBA system, where the chemical structures of ICBA (C$_{78}$H$_{16}$) and PC$_{70}$BM (C$_{82}$H$_{14}$O$_2$) are very similar. In contrast, since in UPS depth profiling, the distinction between materials is based on differences in their electronic structure, which varies greatly between ICBA and PC$_{70}$BM (**Fig. 3**, top), these two chemically similar materials can be easily differentiated. Consequently, while XPS depth profiling (**Fig. 5b, top**) struggles to resolve the different fullerne:fullerene mixing ratios, UPS depth profiling (**Fig. 5b, bottom**) provides far better results. Furthermore, UPS depth profiling reveals the formation of a thin PBDB-T-2Cl overlayer, most likely formed due to the lower surface energy of the donor polymer,[46] which is not observed using XPS depth profiling due to its large probing depth. A similar overlayer is also observed in the case of PTB7, although in this case we also observe that approximately the top 20 nm of the ternary blend are donor-rich. These results confirm the efficacy of UPS depth profiling in characterizing the vertical stratification in ternary organic blends which is critically important for the understanding of the photovoltaic device performance.

## (a) PBDB-T-2Cl:PC$_{70}$BM:ITIC-2F

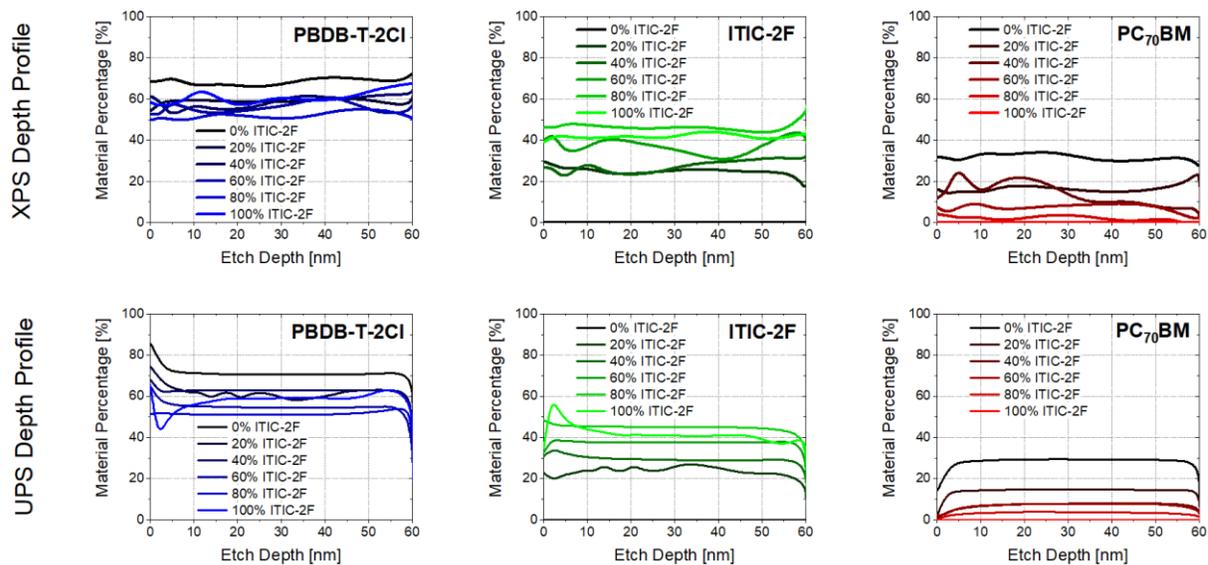

## (b) PTB7:PC$_{70}$BM:ICBA

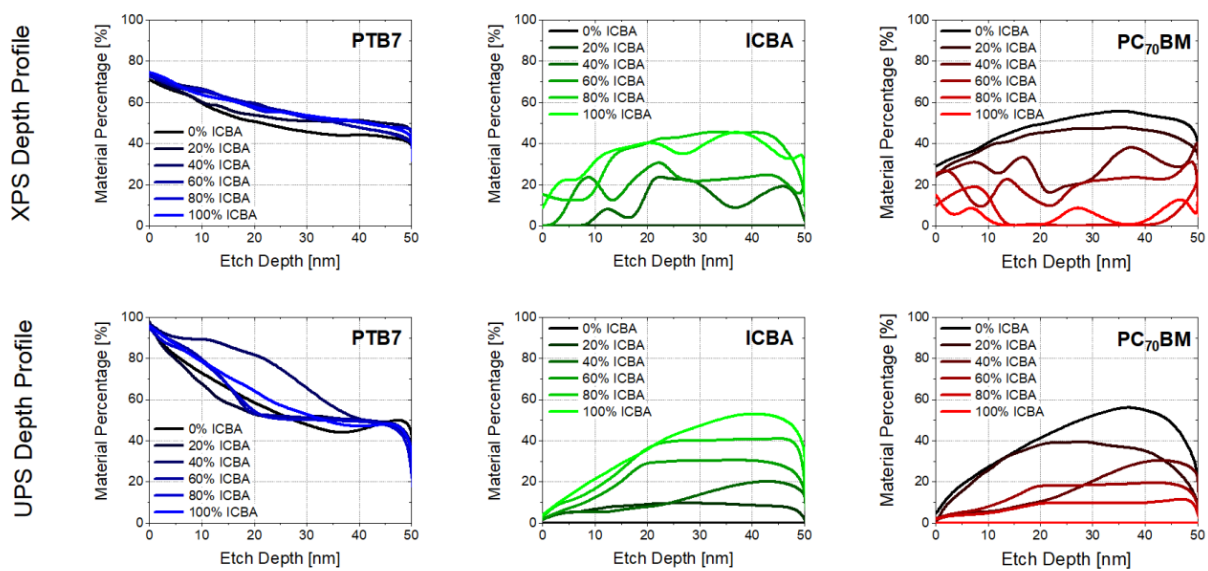

**Figure 5**. Vertical composition profiles obtained via XPS and UPS depth profiles for (a) PBDB-T-2Cl:PC$_{70}$BM:ITIC-2F and (b) PTB7:PC$_{70}$BM:ICBA ternary systems for various ITIC-2F and ICBA weight ratios, respectively.

## CONCLUSIONS

In summary, we investigated the energetic alignment of two ternary systems with both non-fullerene and fullerene acceptors. We demonstrate that energy level diagrams obtained by UPS measurements of the individual materials do not represent the energetic landscape of the ternary system. On the other hand, UPS depth profiling allows for an accurate estimation of the photovoltaic gap of the ternary blend, which shows an excellent agreement with the measured device $V_{OC}$ for all acceptor:acceptor mixing ratios. These results suggest that the sum of radiative and non-radiative energetic losses in the investigated ternary systems is independent of the material mixing ratio and is not the cause of $V_{OC}$ tuning. In addition, we have shown the superiority of UPS over XPS depth profiling in resolving compositional profiles for material combinations with very similar chemical, but dissimilar electronic structures. We believe that the UPS depth profiling method will help to further the understanding of complex ternary or even quaternary systems by providing accurate energetic and compositional information necessary for the design and optimization of future generations of multi-component photovoltaic devices.

**EXPERIMENTAL SECTION**

**Device Fabrication**

Pre-patterned ITO-coated glass substrates (PsiOTec Ltd, UK) were subsequently sonicated in acetone and in isopropanol for 5 min each and then treated with oxygen plasma for 10 min. Immediately after cleaning, a ZnO sol-gel was spin coated at 2000 rpm for 45 s onto the ITO and annealed for 30 min at 200 °C in air.[47,48] Next, the samples were transferred to a nitrogen glovebox where the active layer was deposited (see Active Layer Fabrication). On top the active layer, 10 nm of $MoO_3$ and 80 nm of Ag were thermally evaporated under high vacuum ($10^{-6}$ mbar) to complete the device.

**Active Layer Fabrication (PBDB-T-2Cl:PC$_{70}$BM:ITIC-2F)**

The active layer was entirely fabricated under nitrogen atmosphere. First, the donor material PBDB-T-2Cl and the acceptor materials PC$_{70}$BM and ITIC-2F were dissolved in chlorobenzene mixed with 1 % 1,8-diiodooctane at a concentration of 10 mg/ml each, then heated overnight at 60 °C. All materials were filtered using a 0.4 µm filter. Next, the filtered donor and the two acceptor solutions were mixed at a total D:A volume ratio of 1:1, therein the PC70BM:ITIC-2F weight ratio was varied from 0 to 100 %. Afterwards, the solution was spin-coated in a two-step process at 1000 rpm for 50 s followed by 2000 rpm for 5 s and the resulting film annealed at 100 °C for 10 min.

**Active Layer Fabrication (PTB7:PC$_{70}$BM:ICBA)**

The active layer was entirely fabricated under nitrogen atmosphere. First, the donor material PTB7 was dissolved at a concentration of 20 mg/ml and the acceptor materials PC$_{70}$BM and ICBA were dissolved at a concentration of 30 mg/ml each, then heated overnight at 60 °C. The solvent for all three materials was 1,2-dichlorobenzene mixed with 3 % 1,8-diiodooctane. All materials were filtered using a 0.4 µm filter. Next, the filtered donor and the two acceptor solutions were mixed at a total D:A volume ratio of 1:1.5, therein the PC$_{70}$BM:ICBA weight ratio was varied from 0 to 100 %. Afterwards, the solution was spin-coated in a two-step process at 1000 rpm for 50 s followed by 2000 rpm for 5 s and the resulting film annealed at 100 °C for 10 min.

**Current Density-Voltage (J-V) Characteristics**

A source-measure unit (Keithley 2450) was used to record current density-voltage curves of the TSC devices illuminated under AM 1.5 conditions (ABET Sun 3000 class AAA solar simulator) in ambient conditions.

**External Quantum Efficiency (EQE)**

EQE was measured with the monochromated light of a halogen lamp (range: 350 nm to 900 nm), calibrated with an NIST-traceable Si diode (Thorlabs), in ambient conditions.

**Ultraviolet Photoemission Spectroscopy (UPS) and X-ray Photoemission Spectroscopy (XPS)**

For UPS and XPS measurements, the samples were transferred to an ultrahigh vacuum chamber (ESCALAB 250Xi, Thermo Scientific) with a base pressure of $2 \times 10^{-10}$ mbar. XPS measurements were carried out using an XR6 monochromated Al Kα source (hυ = 1486.6 eV) and pass energy of 20 eV and a spot size of 650 μm. UPS measurements were performed using a He discharge lamp (hυ = 21.2 eV), a pass energy of 2 eV and a sample bias of -5 V. For UPS of the individual materials, pure films were prepared following the same procedures described in the device fabrication, but without mixing materials and evaporating silver contacts.

**Depth Profiling**

UPS and XPS depth profiling was carried out with the MAGCIS dual mode ion source (ESCALAB 250Xi, Thermo Scientific) operated in the argon gas cluster ion beam mode, which has been shown to minimize etching induced material damage.[44,45] The cluster mode was used at an energy of 4 keV with "large" clusters (in the order of $Ar_{2000}$) and an etch crater size of $(2.5 \times 2.5)$ mm$^2$. To create a depth profile, XPS or UPS measurements were altered with cluster etch steps and a 10s break after etching to prevent charging. Depth profiling experiments were performed on the photovoltaic devices, in the areas between the Ag contacts to ensure comparability to the measured $V_{OC}$. Corresponding UPS and XPS depth profiles of a specific material were always measured on the same sample.

**AUTHOR INFORMATION**

**Funding**

This project has received funding from the European Research Council (ERC) under the European Union's Horizon 2020 research and innovation programme (ERC Grant Agreement n° 714067, ENERGYMAPS).

**Notes/Conflict of Interests**

The authors declare no competing financial or commercial interests.

**ACKNOWLEDGMENTS**

The authors would like to kindly thank Prof. U. Bunz for providing access to film fabrication facilities.